# We cannot believe we overlooked these Majorana discoveries

Sergey Frolov, University of Pittsburgh
Vincent Mourik, FZ Juelich

**Abstract.** In 2011-2012 we performed experiments on hybrid superconductor-semiconductor nanowire devices which yielded signatures of Majorana fermions based on zero-bias peaks in tunneling measurements. The research field that grew out of those findings and other contemporary works has advanced significantly, and a lot of new knowledge and insights were gained. However, key smoking gun evidence of Majorana is still lacking. In this paper, we report that while reviewing our old data recently, armed with a decade of knowledge, we realized that back in 2012 our results contained two breakthrough Majorana discoveries. Specifically, we have observed quantized zero-bias peaks, the hallmark of ideal Majorana states. Furthermore, we have observed the closing and re-opening of the induced gap perfectly correlated with the emergence of the zero-bias peak - clear evidence of the topological quantum phase superconducting transition. These insights should pave the way to topological Majorana qubits, and you should also check supplementary information for important disclosures.

Mar 17, 2022

## Introduction

Majorana fermions are at the convergence of the most important tendencies in fundamental physics. The past and the future: embodied by the century-long link between the original work of Ettore Majorana in 1930's [1] and the bright fault-tolerant quantum computing prospects of 2030's. Of particle physics, where the notion is famous through neutrinos, as well as of condensed matter physics, where it is associated with quasiparticles. Astrophysics is also in this mix through connections to dark matter and the idea to emulate the chaos within black holes with a Majorana mode soup. The most profound aspect of this science is the possibility that these objects would demonstrate non-Abelian exchange rules, breaking with all known particles in the universe.

It does not come as a surprise that, with so much riding on these effects existing, the work put into their discovery has been multi-pronged and extensive. Focusing on condensed matter systems, Majorana modes have been proposed and searched for in dozens of different materials and substances, some as elusive as the modes themselves, e.g. quantum spin liquids or p-wave superconductors.

In contrast with these exotic options, a visible segment of the field has focused on very conventional materials, namely on semiconductors and superconducting metals. These are well-studied and found in consumer electronic chips, and in the more experimental but robust quantum logic circuits. What is required for generating Majorana this way, according to theory, is a combination of a standard superconductor with a low-dimensional semiconductor such as a nanowire.

Our own work from 2012 [2], and several contemporary works, have kick-started the experimental effort based on InSb or InAs nanowires coupled to metals such as Nb and Al. We focused on the theoretically predicted signature of Majorana fermions, a peak in conductance (which is a derivative of current with respect to voltage, an inverse of resistance) located at zero source-drain bias in electrical current-voltage transport measurements. By passing a current through a nanowire that was connected to a superconductor, we observed such peaks and verified that they satisfy several criteria relevant for the Majorana hypothesis, such as being present at large magnetic fields.

While the evidence was consistent with Majorana, it was not sufficiently conclusive, and only provided initial signatures of Majorana modes in the nanowire platform. Subsequently, more signatures of the same effect were obtained. The quality of nanowire devices has improved and effects such as 'hard induced gap' and 'high mobility' were reported, both advantageous for putative quantum computing applications of Majorana physics. However, despite all the progress made, the ultimate and conclusive evidence of Majorana modes has remained out of reach.

In this paper, we report clear and unexpected results that put an end to the debate and demonstrate unambiguously the presence of Majorana modes in nanowire devices. The unexpected aspect of our findings is that the evidence was uncovered in our original data from 2012. The advancement of open science principles has recently compelled us to share all of the data from that experiment on Zenodo [3]. Looking through these data, we found two previously overlooked pieces of evidence that complete the Majorana puzzle.

First, we now report on the observation of zero-bias conductance peaks of the expected quantized value of $2e^2/h$. Quantized peaks have been predicted by Majorana theory going back to 2001. Yet, in our own data from 2012 we have reported peaks of much smaller value of 5% of the quantized value. The reason for that was that at the time we focused on other aspects such as peak width and did not appreciate the importance of peak height. Having been further educated in Majorana physics, we now recognize that we have overlooked this major discovery, which we report after a 10-year delay.

Second, Majorana theory is based squarely around the idea that Majorana modes emerge upon a quantum phase transition into the topologically superconducting state. While zero-bias peaks themselves may be taken as a manifestation of such a transition, the smoking gun evidence must come from the closing and subsequent reopening of the superconducting gap, a fact that can be verified with any condensed matter theorist. This effect has never been reported. Upon re-examination of the 2012 data, we have presently identified such a gap closing and reopening. The reopening is perfectly correlated with the appearance of the zero-bias peak on the side of the topological transition.

These two additional and overlooked discoveries, taken together with the signatures reported in 2012, allow us to conclusively establish the existence of Majorana fermions in nanowires. These findings conclude the physics part of the work and open doors to the quantum engineering of Majorana qubits.

The paper is organized as follows. First, we present the first major discovery. Second, we present the second major discovery. Be sure to also carefully read the supplementary information.

**Figure 1 - Quantized Zero Bias Peak**

When our 2012 data appeared in Science in 2012 [2] it was closely analyzed by a large number of high level experts who produced their own papers, both theoretical and experimental, to interpret our findings. As was common at the time, only a limited amount of data was included with the publication, of order 30 data sets. The way the zero-bias peaks looked, what they were surrounded by, how they evolved when experimental parameters were varied in those 30 datasets was given very large significance. The interpretation of the features went beyond what the data were meant to show, since we as key experimental authors were mostly trying to present samples of behavior, not universal features.

Some aspects of the data were more representative than other aspects. Now, with nearly 4000 datasets available on Zenodo [3], we can see that there were many examples of zero-bias peak conductance values much higher than 5% of the quantized value that was presented by us in the main text. At the time, several theories appeared that explained why exactly 5% was expected for Majorana. Many more papers quoted this relatively low value as a reason why experiments need to be improved, in justification of their own work. However, these conclusions went too far in terms of interpreting our results. Even in the supplementary information of Science-2012 there were peaks of the value up to $0.6 \times 2e^2/h$ [2], which went unnoticed.

But the findings we are presenting now are from a qualitatively different class. For the first time, we demonstrate that in 2012 devices, quantization of conductance has been directly demonstrated. Quantized conductance is a phenomenon familiar from quantum point contacts and quantum Hall effects. In the context of Majorana, the origins of it are different. It arises due to the perfect symmetry between tunneling of an electron into the superconductor, through the Majorana mode, and the reverse tunneling of a quasi-hole from the superconductor. This process is also known as resonant Andreev reflection. The idea is so simple and elegant in theory that it must be true.

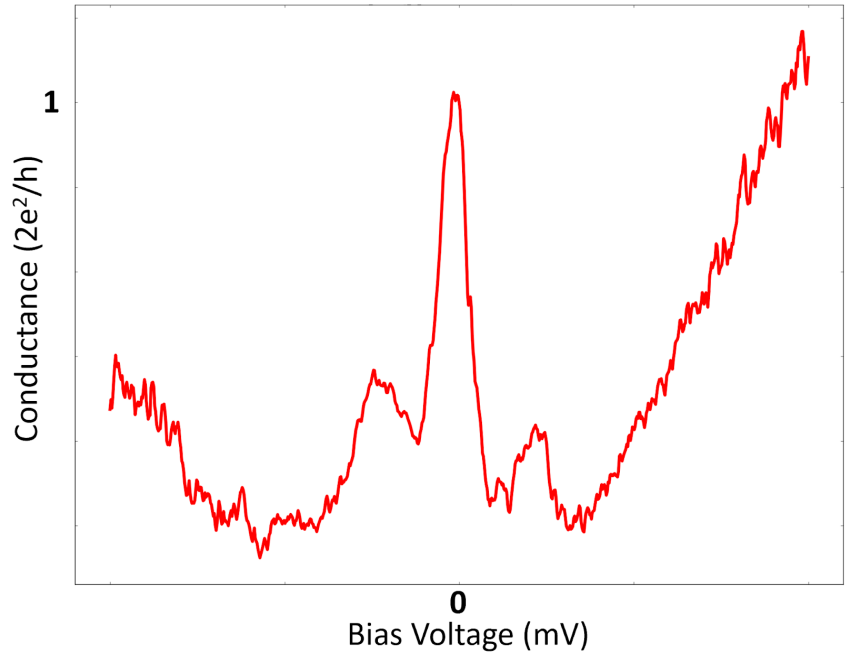

**Figure 1.** *Quantized zero-bias peak is located at bias voltage equal to zero and reaches a conductance value of 1.003 x $2e^2/h$ in a hybrid nanowire device [3].*

Indeed, this is what the data show. At zero source-drain bias, we have observed a peak in conductance that at the tip of the peak reaches the value of 2e2/h, to within 0.3%. The peak is surrounded by two other peaks, which represent the superconducting gap. But the middle peak is the most prominent. The value of conductance speaks for itself. Since it is quantized, the only option for it is due to Majorana. This is because a Majorana peak is expected to be quantized, further confirming that the observed peaks are of a Majorana origin.

**Figure 2 - Closing and Re-Opening of the Topological Gap**

Majorana modes in condensed matter systems are "topological", but this is used in several overlapping senses. On the one hand, they represent an electron split in half and placed on two opposing ends of a nanowire. This provides topological protection to a quantum state of that electron. On the other hand, Majorana modes are a property of topological superconductivity that is also characterized by a topological gap which opens at a topological quantum phase transition. At the transition point, the non-topological (trivial) gap in the spectrum is supposed to close, and the new, topological gap, must re-open. Only after that true topological Majorana modes can appear in the nanowire.

A gap shows up in the data as reduced current or conductance, because within the gap there are fewer or no states to pass a current through. A gap closing would be an increase in conductance when magnetic field or gate voltage (i.e. electric field) is varied. At a phase transition point, conductance would first increase and then decrease. This would happen around zero source-drain bias where superconductivity is observed. Zero bias is also where Majorana modes appear - therefore on one side of the topological phase transition we should be observing Majorana zero modes. This topological phase transition can be driven by experimental parameters that are capable of altering the balance of energies in the sample. Examples of such experimental control knobs are electric fields (gate voltages) and magnetic fields.

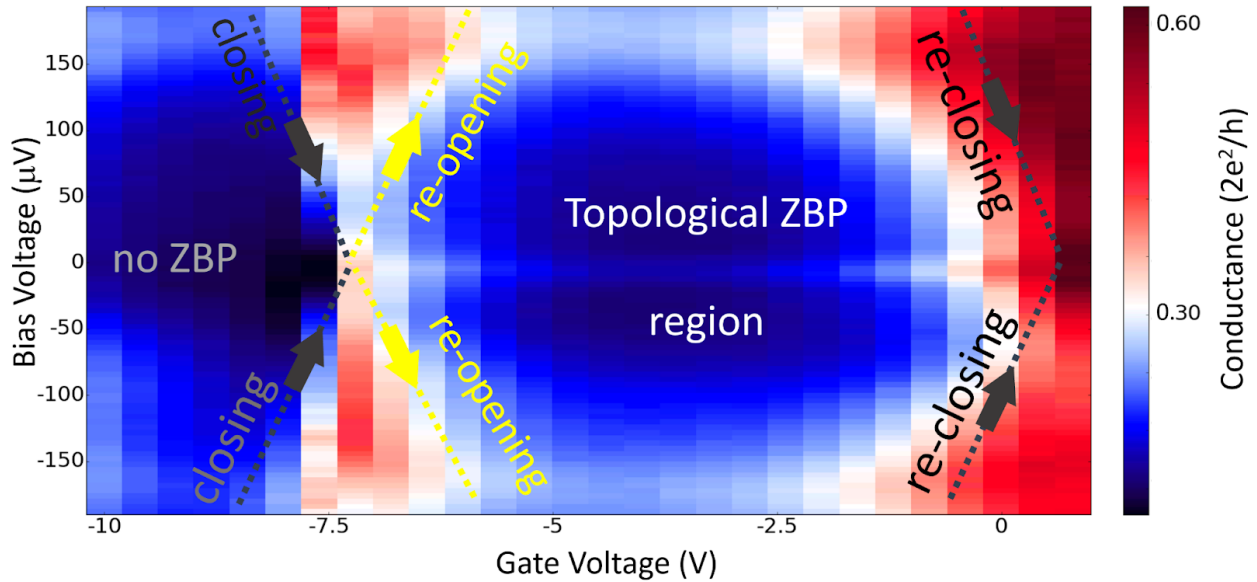

**Figure 2.** *Experimental demonstration of closing and re-opening of the superconducting gap (marked in the figure with lines and arrows). On one side of the topological gap closing, no zero-bias peak (ZBP) is observed because it is the trivial regime. On the other side, for Gate Voltage greater than -7.5 V, a topological zero bias peak appears exactly at the gap re-opening point. As a bonus, gap re-closing is also reported here around zero gate voltage.*

Figure 2 demonstrates the topological phase transition as a function of gate voltage. At the most negative gate voltages, on the left side of the figure, we observed an empty superconducting gap of order 0.2 mV in magnitude (dark-blue region). The phase transition point is at a gate voltage of -7.5V, where we see conductance rising within the gap. This takes shape of a theoretically predicted feature that moves through the gapped region towards zero bias. Upon reaching zero bias voltage, conductance begins to lower again throughout the gap, the gap 're-opens'. Precisely at the quantum phase transition point, the zero-bias peak appears, as expected from the Majorana theory of topological superconductivity. The gap closes for the second time (re-closes) at positive gate voltage around 0 V. At the precise point the zero-bias peak disappears - being replaced by a dip - indicating that the system has transitioned back into the trivial non-topological state, a reverse quantum phase transition.

It is difficult to overstate the significance of this discovery. Even though several reliable and comprehensive numerical models do demonstrate that in realistic devices a Majorana zero-bias peak does not need to be accompanied by the visible closing and re-opening of the gap, it will never be really a Majorana without such a phase transition being demonstrated. An expert would notice that data in Figure 2 are somewhat limited and in low resolution especially in gate voltage. Even though the phase transition is clearly seen, and the coincidence of the gap closing feature and zero-bias peak cannot come from anything other than Majorana physics, we have simply not recognized how consequential this regime was back in 2012, and have not taken more data at the time.

**Discussion**

While in Figure 2 the peak is not quantized, due to trivial effects that are easily explained, the unmistakable pattern of a phase transition, and the coincidence of gap re-opening with the appearance of the zero-bias peak, is very strong evidence of a topological quantum phase transition. In order to stay fully rigorous, we do not claim the observation of Majorana based on just this one piece of data in Figure 2. However, given that in Figure 1 we do observe a clearly quantized peak, from the combination of the two discoveries, we can now definitively conclude that within the 10-year old Majorana data there has been complete and definitive evidence of Majorana modes. This has now been uncovered and properly discussed.

There is not much left to do on the physics side in terms of demonstrating Majorana, at least in nanowires. Perhaps it would be good to make sure that these discoveries can be repeated in newer samples, but this is not expected to be a big issue. Future steps for the field involve braiding of Majorana modes, which can serve as an extra check of the theory, though we conclude that it has now been proven beyond reasonable doubt. Braiding is still a worthy goal given applications in quantum computing that have been associated with it.

**References**

1. E. Majorana, Teoria simmetrica dell'elettrone e del positrone, Nuovo Cimento 14, 171 (1937). English translation
2. V. Mourik, K. Zuo, S.M. Frolov, S.R. Plissard, E.P.A.M. Bakkers, L.P. Kouwenhoven, Signatures of Majorana Fermions in Hybrid Superconductor-Semiconductor Nanowire Devices, Science 336, 6084, 1003 (2012). arXiv:1204.2792
3. Vincent Mourik, Kun Zuo, Sergey Frolov, Sebastien Plissard, Erik Bakkers, & Leo Kouwenhoven. (2021). Full data for 'Signatures of Majorana fermions in hybrid superconductor-semiconductor nanowire devices' Science 336, 1003-1007 (2012) [Data set]. In Science (Vol. 336, pp. 1003–1007). Zenodo. https://doi.org/10.5281/zenodo.5106663
4. H. Zhang, C.-X. Liu, S. Gazibegovic, D. Xu, J. A. Logan, G. Wang, N. van Loo, J. D. S. Bommer, M. W. A. de Moor, D. Car, R. L. M. Op het Veld, P. J. van Veldhoven, S. Koelling, M. A. Verheijen, M. Pendharkar, D. J. Pennachio, B. Shojaei, J. S. Lee, C. J. Palmstrøm, E. P. A. M. Bakkers, S. Das Sarma, L. P. Kouwenhoven, RETRACTED ARTICLE: Quantized Majorana conductance, Nature volume 556, 74 (2018). arXiv:1710.10701
5. S. Vaitiekenas, G. W. Winkler, B. van Heck, T. Karzig, M.-T. Deng, K. Flensberg, L.I. Glazman, C. Nayak, P. Krogstrup, C. M. Marcus, Flux-induced topological

superconductivity in full-shell nanowires, Science, 367, 6485 (2020). arXiv:2003.13177

**Supplementary Information**

Claims made in the main text of this article are not reliable. They are a result of extremely narrow (non-representative) data selection. This means that, while the data used to create Figures 1 and 2 are real, they are not representative of the experiments. It is for this reason that they were not emphasized this way within the initial report of these results in 2012. In short, the zero-bias peak height is highly sensitive to all experimental parameters, and a given value of the peak is not meaningful in itself even if it does find itself close to a quantized value like in Figure 1. (And of course the value of 5% is also not significant). The emergence of a zero-bias peak next to a feature transitioning through a gap, like in Figure 2, is a pure coincidence.

The purpose of this article is to illustrate how profound the role of data selection can be in condensed matter physics publications. If a reader thinks that examples chosen were too obvious, we would like to bring to their attention that exactly these arguments were used to make Majorana claims in a recent Nature paper [4], now retracted, and in a recent Science paper [5], now under the editorial expression of concern.

Many more examples of biased data selection, or cherry picking, exist in our field. Indeed, a large fraction of papers our community produces are just figures narrated by text. Meaning, the figures show a one-off data acquisition. An experimentalist pushed a button on their measurement system and obtained a single picture. They have done it multiple times but they only show a single image in the paper. This is similar to a photographer picking a photo for their portfolio out of a large series of images.

We have a habit of trusting the author in that they chose a representative picture for us to look at, and we look no further. Especially if the picture shows a pattern that we already expect, be it from theory or from other papers.

Below we show how we have performed cherry-picking for Figures 1 and 2 of this paper. In Figure S1 we demonstrate how, as a function of the magnetic field, we can find a vertical trace in which the zero-bias peak has a high value, but is surrounded by lower peaks or traces with no peaks at all. On the other hand, it could be argued that the peak forms a 'plateau' and that it 'never exceeds that quantized value', which are all good things to say if one wants to argue for Majorana. The plateau effect can be attained by increasing the magnetic field resolution and subsequently stretching the magnetic field region where peaks reach the right value.

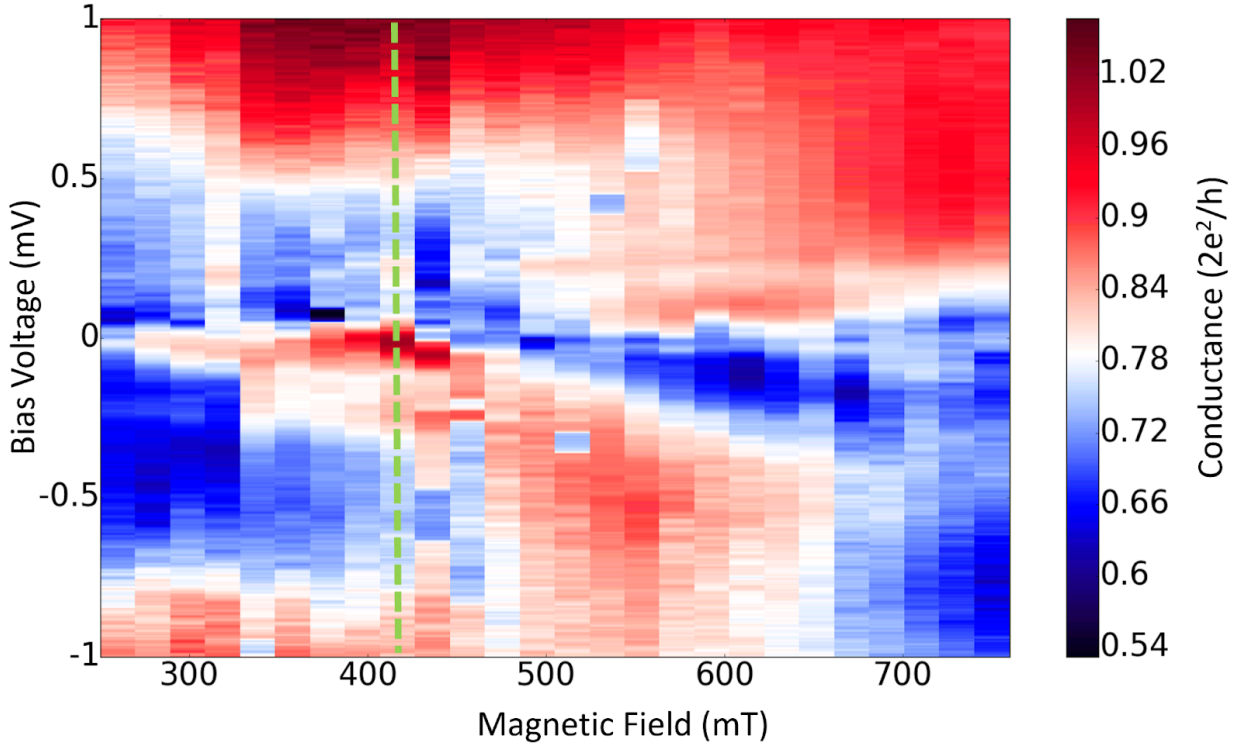

**Figure S1.** *A full dataset from which Figure 1 of the main text was derived. Green dashed line marks the magnetic field at which the data for Figure 1 originate. It is surrounded by regions of zero-bias peaks, of varied heights, as well as regions without zero-bias peaks.*

To illustrate that Figure 2, the 'topological phase transition' is just a coincidence, we show the data in Figure S2, which is qualitatively similar to Figure 2. Indeed we see a gap feature at a bias of 0.2 mV, and at a gate voltage near -5V we see a feature crossing the gap which we described as a gap closing and re-opening. There is a zero-bias peak to the right of this feature. But, in contrast to Figure 2, in Figure S2 there is also a zero-bias peak to the left of the feature. This is not expected from a Majorana topological phase transition. Even though the peak itself may be of Majorana origin, the X-shaped artefact at the gate voltage of -5V does not signify the closing and re-opening of the gap. Its likely origin is a quantum dot state in the nanowire.

We also included a larger voltage bias range (vertical axis) for this figure. This makes it clear that the gap feature at constant bias is always present, while the criss-cross feature exists both below and above the gap. Thus, there isn't any gap closing. This is also the case for the main text Figure 2, but there we reduced the range of the vertical axis to conceal the fact that the criss-cross feature is present above the gap.

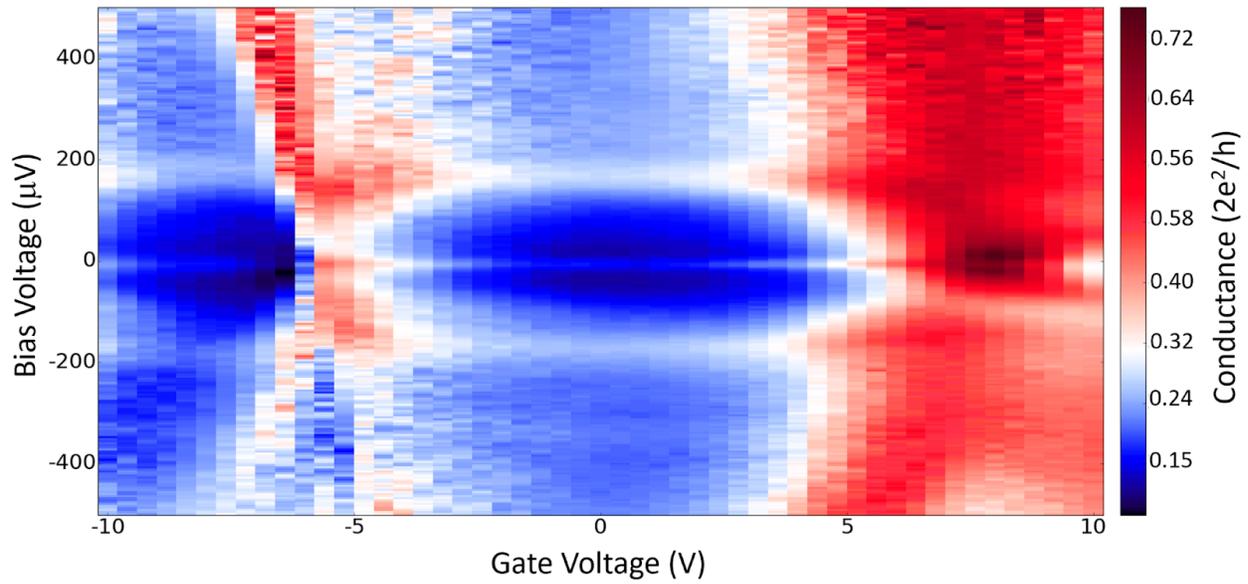

**Figure S2.** *A full and uncropped dataset analogous to the data in the main text Figure 2. Here, zero-bias peak exists on either side of the criss-crossing feature at the gate voltage of -5 V. The feature itself extends above the gap thus cannot be indicative of the gap closing.*

In a recent Science paper from 2020, Figure 2 shows a coincidence of zero-bias peaks and another dramatic effect - Little-Parks oscillations that lead to actual quasiperiodic closing of the gap [5]. Based on this coincidence, that paper makes a claim that zero-bias peaks are due to Majorana modes driven by magnetic flux that also induces these Little-Parks oscillations. Additional data from those experiments show that Figure 2 to be just a coincidence, and in other data the zero-bias peak does not coincide with Little-Parks oscillations.

Demonstration of Majorana modes in condensed matter systems remains an interesting, important and open challenge. The purpose of this paper is to increase awareness of non-representative data selection issues within this topic, and in condensed matter physics at large. We hope these problems can be avoided in the future. This should increase the reliability of reported science and, in the longer run, accelerate progress in the field.